\documentclass[11pt]{article}
\usepackage{amsmath,amssymb,graphicx}

\newcommand{\ee}{e}
\newcommand{\im}{i}
\newcommand{\spc}{{\,}}
\newcommand{\ft}[2]{{\textstyle\frac{#1}{#2}}}
\newcommand{\gtop}{g_{\mathrm{top}}}
\newcommand{\Sc}{\mathcal{S}}
\newcommand{\Fc}{\mathcal{F}}
\newcommand{\Ftop}{F_{\mathrm{top}}}
\newcommand{\Ztop}{Z_{\mathrm{top}}}
\newcommand{\Yf}{Y^0_{\mathrm{f}}}
\newcommand{\abs}[1]{\lvert #1\rvert}
\renewcommand{\Im}{\operatorname{Im}}
\renewcommand{\Re}{\operatorname{Re}}
\DeclareMathOperator{\Li}{Li}
\newcommand{\pd}{\partial}
\newcommand{\bigO}{\mathcal{O}}

\numberwithin{equation}{section}
\begin{document}
\begin{titlepage}
\begin{center}
\hfill LMU-ASC 15/06\\
\hfill MPP-2006-25\\
\hfill {\tt hep-th/0603211}\\
\vskip 10mm

{\Large \textbf{Entropy Maximization in the Presence of \\\vskip 2mm
Higher-Curvature Interactions }}
\vskip 8mm

\textbf{G.~L.~Cardoso$^{a}$,  D.~L\"ust$^{a,b}$ and J.~Perz$^{a,b}$}

\vskip 4mm
$^{a}${\em Arnold Sommerfeld Center for Theoretical Physics\\
Department f\"ur Physik,
Ludwig-Maximilians-Universit\"at M\"unchen \\
Theresienstra{\ss}e 37,
80333 M\"unchen, Germany}\\

\vskip 4mm
$^{b}${\em Max-Planck-Institut f\"ur Physik \\
F\"ohringer Ring 6,
80805 M\"unchen, Germany}\\
\vskip 4mm

{\tt gcardoso,luest,perz@theorie.physik.uni-muenchen.de}
\end{center}
\vskip .2in
\begin{center} {\bf ABSTRACT } \end{center}
\begin{quotation}\noindent
  Within the context of the entropic principle,
  we consider the entropy of supersymmetric black holes in $N=2$
  supergravity theories in four dimensions with higher-curvature
  interactions, and we discuss its
  maximization at points in moduli
  space at which an excess of hypermultiplets becomes massless.  We
  find that the gravitational coupling function $F^{(1)}$ enhances the
  maximization at these points in moduli space.  In principle, this
  enhancement may be modified by the contribution from higher
  $F^{(g)}$-couplings.  We show that this is indeed the case for the
  resolved conifold by resorting to the non-perturbative expression for
  the topological free energy.
\end{quotation}

\vfill
\end{titlepage}

\eject

\section{Introduction}

In four dimensions, the near horizon geometry of BPS black hole
solutions is characterized by attractor equations
\cite{Ferrara:1995ih,Strominger:1996kf,Ferrara:1996dd} which, at the
two-derivative level, follow from the extremization condition of the
black hole central charge $Z$, i.e.~$D Z = 0$.  The latter exhibits
an interesting similarity to the condition $D W=0$ for supersymmetric
flux vacua, where $W$ denotes the flux-generated superpotential in
type II or F-theory compactifications.  A connection
\cite{Ooguri:2005vr} between black holes and flux compactifications is
provided by type IIB BPS black hole solutions, for which the near
horizon condition $D Z = 0$ can be viewed as the extremization
condition of a five-form flux superpotential $W$ generated upon
compactifying type IIB string theory on $S^2 \times M$, where $M$
denotes a Calabi--Yau threefold (for related work see
\cite{Curio:2000sc,Behrndt:2001mx,Denef:2004ze,Kallosh:2005bj}).  The
resulting $1+1$-dimensional space-time has a negative cosmological
constant determined by the value of $W$ at the extremum
\cite{Ooguri:2005vr}.

In view of this connection, it was suggested in
\cite{Ooguri:2005vr,Gukov:2005bg} to interpret the
exponentiated entropy of large
BPS black holes in Calabi--Yau compactifications as an entropic
function for supersymmetric flux compactifications on $AdS_2 \times
S^2 \times M$.  At the two-derivative level, the entropy of a black
hole is given by the area law of Bekenstein and Hawking, which for
large BPS black holes takes the form \cite{Behrndt:1996jn}
\begin{equation}
  \Sc = \pi \, \abs{Y^0}^2 \, \ee^{- G (z, \bar z)} \spc,
\label{entropy2d}
\end{equation}
where, in a certain gauge, $G (z, \bar z)$ reduces to the
K\"ahler potential for the moduli fields $z^A = Y^A/Y^0$ belonging to
vector multiplets labelled by $A = 1 , \dotsc, n$.  The fields $Y^0$
and $z^A$ are expressed in terms of the black hole charges $(p^I,q_I)$
by the attractor equations.  Once the charges are identified with
fluxes,
each choice $(Y^0, z^A)$
translates into a particular flux compactification. By fixing $Y^0$
to a specific value $\Yf$, the entropy (\ref{entropy2d})
can be viewed as a function over the moduli space of the
Calabi--Yau threefold, and to each point $z^A$ in moduli space one
assigns a (suitably normalized) probability density $\ee^\Sc$
(entropic principle \cite{Ooguri:2005vr,Gukov:2005bg}).

The entropy of BPS black holes is corrected by higher-curvature
interactions \cite{LopesCardoso:1998wt}.  Therefore, the probability
density for $AdS_2$ vacua with five-form fluxes is modified due to
$R^2$-interactions.  In this paper, we will be interested in studying
the maximization of the entropy, viewed as a function over the
moduli space of the Calabi--Yau threefold, in the presence of
higher-curvature corrections.

We study the entropy extremization in the neighborhood of certain
singularities of Calabi--Yau threefolds. Our results differ from
\cite{Gukov:2005bg}, because
in contrast to \cite{Gukov:2005bg},
we do not consider the extremization of the Hartle--Hawking type wave
function
$|\psi_{0,0}|^2$,\footnote{We thank S.~Gukov, K.~Saraikin and C.~Vafa
for clarification of this point.}
but instead the extremization of the
wave function $|\psi_{p,q}|^2$,
whose value at the attractor point is
the exponential of the entropy
\cite{Ooguri:2005vr}.
We find that
singularities where an excess of additional massless hypermultiplets
appear, correspond to local maxima of the entropy. Therefore,
following the entropic principle,
the associated vacua would have a higher probability.

We demonstrate that the gravitational coupling $F^{(1)}$ leads to an
enhancement of the maximization of the entropy at these singularities.
For the case of the conifold, we also take into account the
contribution from the higher coupling functions $F^{(g)}$ by resorting
to the non-perturbative expression of the topological free energy
computed in \cite{Gopakumar:1998ii,Gopakumar:1998ki} for the resolved
conifold.  We find that the entropy is maximized at the conifold point
for the case of real topological string coupling constant, whereas it
ceases to have a maximum at the conifold point for complex values of the
coupling constant.

This paper is organized as follows.  In section \ref{entrofun} we
discuss the
entropy as a function on the moduli space of Calabi--Yau
compactifications.  Since the number of physical moduli $z^A$ is one
less than the number of pairs of black hole charges $(p^I,q_I)$, one
has to fix one particular charge combination in order to discuss the
maximization of the entropy with respect to the $z^A$.  One way to do
this is to set $Y^0$ to a constant value throughout moduli space,
which implies that the topological string coupling constant is held
fixed.  This is reviewed in section \ref{fixy0}.
In section \ref{emwr2} we discuss the
maximization of the entropy computed from the prepotential $F^{(0)}$.
We give a basic example which shows that entropy maximization occurs
whenever a surplus of hypermultiplets becomes massless at the
singularity.  We then comment on various concrete models.  In section
\ref{emr2} we discuss entropy maximization in the presence of
higher-curvature
interactions by using the genus expansion of the topological free
energy.  For the case of the resolved conifold, we also use the
non-perturbative expression for the topological free energy to discuss
entropy maximization near the resolved conifold singularity.
In section  \ref{osvfree} we discuss the minimization of the OSV free
energy \cite{Ooguri:2004zv}.  Section
\ref{concl} contains our conclusions, and appendix \ref{normalztn}
our normalization
conventions.

\section{The entropic function \label{entrofun}}

We begin by recalling various properties of the entropy of
four-dimensional BPS black holes in $N=2$ supergravity theories.  In
the absence of higher-curvature corrections, the entropy is given by
the area law of Bekenstein and Hawking, which for BPS black holes
takes the form \cite{Behrndt:1996jn}
\begin{equation}
  \Sc = \pi \,\im\left(
  \bar Y^I \, F_I^{(0)} (Y) - Y^I \, \bar F_I^{(0)} (\bar Y) \right)
  = \pi \, \abs{Y^0}^2 \, \ee^{- G(z, \bar z)} \spc.
\label{entro2d}
\end{equation}
The fields $Y^I$ are
determined in terms of the charges carried by the black hole
by virtue of the attractor
equations (see section \ref{fixy0}).
Here,
\begin{equation}
F^{(0)} (Y) = -\im \, (Y^0)^2 \, \Fc^{(0)} (z) \spc, \qquad
z^A = Y^A/Y^0 \spc,
\label{f0f0}
\end{equation}
denotes a holomorphic
function which is homogeneous of degree two, i.e.~$F^{(0)} (\lambda Y) =
\lambda^2 F^{(0)} (Y)$ for any $\lambda\in\mathbb{C}\setminus\{0\}$.
The indices run over $I = 0, \dotsc, n$ and
$A = 1, \dotsc, n$, and $ F_I^{(0)} = \partial F^{(0)}/\partial Y^I$.
The quantity $G (z, \bar z)$ is given by
\begin{equation}
\ee^{-G (z, \bar z)} =
2 \left(
\Fc^{(0)} + \bar\Fc^{(0)} \right)
 - (z^A - \bar z^A) \left( \Fc_A^{(0)}
 - \bar\Fc_A^{(0)} \right) ,
\label{G2d}
\end{equation}
where $\Fc_A^{(0)} = \partial\Fc^{(0)}/\partial z^A$.
The $Y^I$ are related to the holomorphic sections $X^I(z)$ of special
K\"ahler geometry \cite{deWit:1983rz,deWit:1984pk,Strominger:1990pd}
by $Y^I = \ee^{K(z , \bar z)/2}\, \bar Z\, X^I (z) $,
where $Z$ denotes the central charge, and where $K$ denotes the K\"ahler
potential
\begin{equation}
K (z, \bar z) = - \log \left(\im \bar X^I(\bar z)
\, F_I^{(0)} (X(z)) - \im
X^I(z) \, \bar F_I^{(0)} (\bar X(\bar z))
\right) .
\label{kaehlerpot}
\end{equation}
Under K\"ahler transformations,
\begin{equation}
X^I(z) \rightarrow \ee^{-f (z)} \, X^I(z) \spc, \quad
K (z , \bar z) \rightarrow K (z , \bar z) + f(z) + \bar f (\bar z)
\spc,\quad
Z \rightarrow \ee^{- \ft12 [f(z) - \bar f(\bar z)]} \, Z \spc.
\end{equation}
It follows that the $Y^I$ are invariant under K\"ahler transformations,
and
so is $G$.  Comparing (\ref{kaehlerpot}) with (\ref{G2d}) yields
\begin{equation}
G (z, \bar z) =  K (z, \bar z) + \log\abs{X^0(z)}^2  \spc.
\end{equation}
In the gauge $X^0(z) = 1$, we have $G = K$.  On the other hand, in the
gauge $X^0(z) = W(X(z))$, where $W(X(z))$ denotes the holomorphic
central
charge $W(X(z)) = \ee^{- K/2} Z$, we have
$G = K + \log\abs{W}^2 = 2 \log\abs{Z}$
\cite{Bellucci:2006ew}.

The exponentiated entropy
of large BPS black holes in Calabi--Yau compactifications
was proposed in \cite{Ooguri:2005vr,Gukov:2005bg}
as a probability density for
supersymmetric flux compactifications
on $AdS_2 \times S^2 \times M$, where $M$ denotes a Calabi--Yau
threefold.
For this to be an unambiguous assignment, however,
$Y^0$ must be fixed.  One possibility would be to allow $Y^0$ to vary
in a prescribed way
as one moves around in moduli space, i.e.~$Y^0 (z, \bar z)$.  A more
economical possibility consists in assigning the same value $\Yf$ to all
points in moduli space, i.e.~$Y^0 = \Yf$ \cite{Gukov:2005bg}.
The fields $Y^0$ and $z^A$ are expressed
in terms of the black hole
charges by the attractor equations
\cite{Ferrara:1995ih,Strominger:1996kf,Ferrara:1996dd}, as
will be reviewed in
section \ref{fixy0}.
These charges are in turn interpreted as
flux data. Therefore, to each particular flux compactification we can
assign
a probability density proportional to $\ee^\Sc|_{\Yf}$.
Fixing $Y^0$ to a particular
$\Yf$ means choosing a codimension one hypersurface in the complex space
of charges (assuming that the charges are continuous), which provides
a mapping between moduli $z^A$ and charges $(p^I, q_I)$.

Having fixed $Y^0= \Yf$, one may look for maxima
of (\ref{entro2d}) in moduli space, i.e.
for maxima of $\ee^{- G (z, \bar z)}$ in a certain domain.
Local extrema in the interior of this domain satisfy
$\partial_A \,
\ee^{-G (z, \bar z)} =0$.
In order to determine the nature of these critical points one can
analyze the definiteness of the matrix of
second derivatives of $\ee^{-G(z, \bar z)}$.
At a critical point,
$\partial_A \partial_{\bar B} \,
\ee^{-G} = - g_{A {\bar B}} \, \ee^{-G}$,
where $g_{A \bar B} =  \partial_A \partial_{\bar B} K$
denotes the K\"ahler metric.  If
 $\Fc^{(0)}_{ABC}$ and $g_{A \bar B}$ are finite, then
$\partial_A \partial_B \,
\ee^{-G} $ vanishes there.
This can be best seen \cite{Bellucci:2006ew}
in the gauge $X^0(z) = W(X(z))$, where the vanishing of
$\partial_A \partial_B \,
\ee^{-G} $ translates into the vanishing of $\partial_A \partial_B \,
\abs{Z}$.  The latter is guaranteed to hold by virtue of special
geometry
 \cite{Ferrara:1997tw}, provided that $\Fc^{(0)}_{ABC} \,
g^{C \bar C}$ is finite.  By direct calculation,
the vanishing of $\partial_A \partial_B \,
\ee^{-G}$ implies
that $(z^C - \bar z^C) \, \Fc^{(0)}_{ABC} $ vanish.
Thus, it follows that
if the metric $g_{A \bar B}$ is positive definite,
the critical point is a maximum of ${\ee}^{-G}$
\cite{Fiol:2006dv,Bellucci:2006ew}.

In Calabi--Yau compactifications, and for large values of the moduli
$z^A$,
 $\Fc^{(0)} (z)$ is a cubic expression
in the $z^A$ and therefore, the entropy (\ref{entro2d}) grows to
infinity as $z^A \rightarrow \infty$.  Hence, in order to study the
maximization of $\ee^{- G (z, \bar z)}$ in a well-posed way, we restrict
ourselves to a finite region in moduli space and ask, whether the
entropy has local maxima in this region.
As we will discuss in section \ref{emwr2}, a class of such points
is provided by singularities
of the Calabi--Yau threefold at which an excess of
(charged) hypermultiplets becomes massless.
Examples thereof are
the conifold of the mirror quintic \cite{Candelas:1990rm} as well as
singularities
associated with the appearance of non-abelian gauge symmetries with
a non-asymptotically free spectrum
\cite{Klemm:1996kv,Katz:1996ht}.

Consider the case when
the singularity
is characterized by a vanishing modulus $V = -\im z^1$ with
$\Fc^{(0)} (z) \sim p(T) - V^2 \, \log V$,
where $p(T)$ denotes a function of the remaining moduli, which are
held fixed.
This results in $\Fc^{(0)}_{VV} \sim - \log V$ as well as
$\Fc^{(0)}_{VVV} \sim - V^{-1}$,
which diverges as $V \to 0$, while
$(z^1 - \bar z^1) \, \Fc^{(0)}_{111} \sim
(V + \bar V) \Fc^{(0)}_{VVV} \sim (1 + \bar V/V)$ remains
finite.
This is thus an example where
$\Fc^{(0)}_{ABC}$ tends to infinity in such a way that
$\partial_A \partial_B \,
\ee^{-G}$ remains finite and non-vanishing at the singularity.
The function $\ee^{-G}$ has an extremum at $V=0$ (see section
\ref{emwr2}).
Since the metric $g_{A \bar B}$
diverges at the singularity, it follows that
$\partial_A \partial_B \,
\ee^{-G}$ is smaller than $\partial_A \partial_{\bar B} \,
\ee^{-G}$.  Since the metric
 $g_{A \bar B}$ is positive definite near the singularity,
the extremum of $\ee^{-G}$ at $V=0$ is a local
maximum.

The maximization of the entropy may be further enhanced when taking
into account higher-curvature corrections.  This will be discussed in
section \ref{emr2}.
In the presence of such corrections, the entropy ceases to be
given by the area law (\ref{entro2d}).
For the case of a certain
class of terms quadratic
in the Riemann tensor encoded in a holomorphic homogeneous function
$F(Y,\Upsilon)$,
the macroscopic entropy, computed from
the associated effective $N = 2$ Wilsonian action using Wald's law
\cite{Wald:1993nt},
is given by \cite{LopesCardoso:1998wt}
\begin{equation}
\Sc = \pi \Bigl(\im \left( \bar Y^I \, F_I (Y, \Upsilon)
- Y^I \, \bar F_I (\bar Y, \bar \Upsilon) \right)
+ 4 \Im \left( \Upsilon \, F_{\Upsilon} \right) \Bigr)\spc,
\label{entro}
\end{equation}
where
$F_I = \partial F/\partial
Y^I$ and $F_{\Upsilon} = \partial F/ \partial \Upsilon$.
Here $\Upsilon$
denotes the square of the (rescaled) graviphoton `field strength', which
takes
the value $\Upsilon = - 64$ at the horizon of the black hole.  The $Y^I$
are determined in terms of the black hole charges by
the attractor equations (\ref{eq:attractor}).
The term
$\pi \,\im \left[ \bar Y^I \, F_I (Y, \Upsilon)
- Y^I \, \bar F_I (\bar Y, \bar \Upsilon) \right] $
describes the $R^2$-corrected area of the black hole, while the term
$4 \pi \Im \left( \Upsilon \, F_{\Upsilon} \right)$
describes the deviation from the area law due to
the presence of higher-curvature interactions.

\section{Fixing $Y^0$\label{fixy0}}

In the presence of higher-curvature corrections encoded in $F(Y,
\Upsilon)$,
the attractor equations determining the near-horizon values of $Y^I$
take the form \cite{LopesCardoso:1998wt}
\begin{equation}
  \label{eq:attractor}
  Y^I-\bar Y^I =\im p^I\spc,\qquad
  F_I(Y,\Upsilon) - \bar F_I(\bar Y,\bar\Upsilon) =\im q_I\spc,
\end{equation}
where $(p^I, q_I)$ denote the magnetic and electric charges of a
BPS black hole, respectively.
Since $F(Y, \Upsilon)$ is
homogeneous of degree two, i.e.~$F(\lambda Y,\lambda^2\Upsilon) =
\lambda^2
F(Y,\Upsilon)$ for any $\lambda\in\mathbb{C}\setminus\{0\}$, then by
Euler's theorem
\begin{eqnarray}
  \label{eq:homog}
  2 \,F -Y^I F_I = 2\,\Upsilon F_\Upsilon \spc,
\end{eqnarray}
and consequently
\begin{equation}
  \label{eq:fzero}
F_0 = \frac{1}{Y^0} \left(2 F - 2 \Upsilon \, F_{\Upsilon}  - z^A\,
\frac{\partial F}{\partial z^A}
\right),
\end{equation}
where $z^A = Y^A/Y^0$.
Using (\ref{eq:attractor}) we compute
\begin{equation}
z^A \pm \bar z^A = \frac{1}{2 \abs{Y^0}^2} \Bigl( \mp \im
\, p^A (Y^0 \mp
\bar Y^0)
\pm (Y^A + \bar Y^A) (Y^0 \pm \bar Y^0) \Bigr)\spc.
\label{zlin}
\end{equation}
As discussed in the previous section, we would like to fix the value of
$Y^0$
to a constant value $\Yf$ throughout moduli space.
Inspection of (\ref{zlin}) suggests to take either
$\Yf = {\bar Y}^0_{\mathrm{f}} $ or $\Yf = - {\bar Y}^0_{\mathrm{f}}$,
since this leads
to a simplification of the expression.
Note, however, that in order to be able to connect four-dimensional
BPS black holes to spinning BPS black holes in five dimensions
\cite{Gaiotto:2005gf,Behrndt:2005he},
both $q_0$ and $p^0$ have to be non-vanishing, which requires taking
$\Yf$ to be complex.

Setting $\Yf = {\bar Y}^0_{\mathrm{f}}$ implies $p^0 =0$.  Then,
(\ref{zlin})
reduces to
\begin{equation}
z^A - {\bar z}^A = \frac{\im}{\Yf}\, p^A \spc,
\label{attractorzmz}
\end{equation}
and the second attractor equation in (\ref{eq:attractor}) gives
\begin{equation}
\frac{\partial F}{\partial z^A}
- \frac{\partial {\bar F}}{\partial {\bar z}^A}
=\im \Yf \, q_A \spc.
\label{attractorfmf}
\end{equation}
The value of $\Yf$ is
determined by the equation $F_0 - {\bar F}_0 =\im q_0$,
and is expressed in terms of $z, {\bar z}$ and $q_0$.  For instance,
when neglecting $R^2$-interactions, and using (\ref{eq:homog}), it
follows that
\begin{equation}
F_0 = -\im Y^0 \left( 2 \Fc^{(0)}(z) - z^A
\Fc^{(0)}_A (z) \right),
\end{equation}
where we set $F = F^{(0)} = -\im (Y^0)^2 \, \Fc^{(0)} (z)$, and where
$\Fc^{(0)}_A = \partial \Fc^{(0)}/\partial z^A$.  Then
\begin{equation}
\Yf = \bar Y^0_\mathrm{f} = - \frac{q_0}{ 2 ( \Fc^{(0)} +
\bar\Fc^{(0)}) - (  z^A
\Fc^{(0)}_A +  {\bar z}^A {\bar \Fc}^{(0)}_A ) } \spc.
\label{yfix}
\end{equation}
For a fixed value $\Yf$, one moves around in moduli space by changing
the
charges $(p^A, q_A)$, which results in a change of $z^A$ according to
(\ref{attractorzmz}) and (\ref{attractorfmf}).  However,
in order to keep
$\Yf$ constant as one varies $z^A$, it follows from (\ref{yfix})
that
one must change $q_0$ in a continuous
fashion.
Since $q_0$ is
quantized, we need to take $q_0$ to be large in order
to be able to treat it as a continuous variable.
Observe that when $\Yf$ is large, a unit change in the charges
$(p^A,q_A)$
corresponds to a semi-continuous change of the $z^A$.

Similarly, choosing $\Yf = - \bar Y^0_\mathrm{f}$ yields $\Yf =\im
p^0/2$ fixed
to a particular value.
Then, (\ref{zlin}) and
(\ref{eq:attractor}) yield
\begin{equation}
\begin{split}
z^A + {\bar z}^A &= 2 \, \frac{p^A}{p^0} \spc, \\
\frac{\partial F}{\partial z^A} + \frac{\partial {\bar F}}{\partial{\bar
z}^A}
  &= - \frac{q_A \, p^0}{2}\spc.
\end{split}
\end{equation}
A choice of charges $(p^A, q_A)$
determines a point $z^A$, and the
remaining equation $F_0 - {\bar F}_0 =\im q_0$ determines
the value of $q_0$.  This value will, generically, not be
an integer, and therefore consistency requires again taking
$q_0$ to be large in order to be able to treat it as a continuous
variable.

Observe that $Y^0$ is related to the topological string coupling $\gtop$
by (see (\ref{dict}))
\begin{equation}
(Y^0)^2 \, \gtop^2 = 4 \pi^2 \;.
\end{equation}
Therefore, we will be interested in taking $Y^0$ to be real
(i.e.~$\gtop$
real) or complex, but not purely imaginary.

\section{Entropy maximization near singularities \label{emwr2}}

\begin{figure}[b]
\centering
\includegraphics[bb=91 3 322 163]{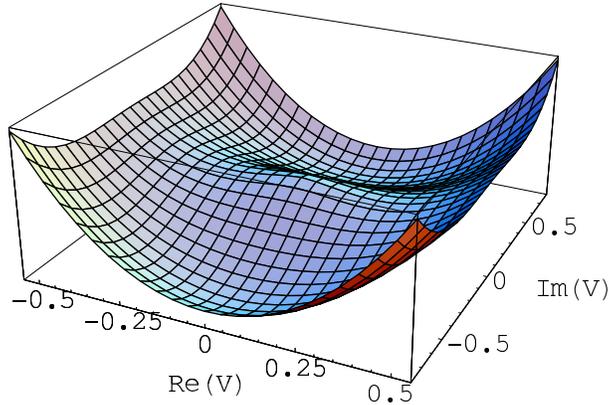}
\caption{$\ee^{- G}$  exhibits a local maximum at $V=0$
for negative $\beta$.\label{gmax}}
\end{figure}

In this section we study the entropy in the absence of higher-curvature
interactions and the maximization of (\ref{G2d})
near singularities of Calabi--Yau threefolds.
Setting $T^A = -\im z^A = -\im Y^A/Y^0$, we obtain
from (\ref{G2d})
\begin{equation}
 \ee^{-G(T, {\bar T})}
=  2 \left(
\Fc^{(0)} + \bar\Fc^{(0)} \right)
 - (T^A + {\bar T}^A) \left( \Fc_{T^A}^{(0)}
 + \bar\Fc_{{\bar T}^A}^{(0)} \right) .
\end{equation}
Let us examine the case when one of the $T^A$ is taken to be small.
We will denote this modulus by $V = -\im z^1 = -\im Y^1/Y^0$.
We consider the situation where
$\ee^{-G}$ is extremized as $V \rightarrow 0$, while $Y^0$
and the remaining moduli $T^a$ are kept fixed.  As our basic example, we
take the following $\Fc^{(0)}$,
 \begin{equation}
  \Fc^{(0)} (V) = \frac{\beta}{2 \pi} \, V^2 \log V + a \spc,
 \label{f0basic}
 \end{equation}
where $\beta$ denotes a real constant, and
where the constant $a$ is complex.  We compute
\begin{equation}
\begin{split}
 \ee^{-G(V, {\bar V})} &= 2 \left(
\Fc^{(0)} + {\bar \Fc}^{(0)} \right)
 - (V + \bar V) \left( \Fc_V^{(0)}
 + {\bar \Fc}_{\bar V}^{(0)} \right) \\
&= 4 \Re a - \frac{\beta}{2 \pi} (V + {\bar V})^2 - \frac{ 2
\beta}{\pi} \abs{V}^2 \log\abs{V} \spc.
 \label{entrot}
\end{split}
\end{equation}
We note that $\ee^{-G(V, {\bar V})}$
has a local maximum at $V=0$ for negative $\beta$,
and that the value at this
maximum is given by
$4 \Re a
\equiv \ee^{-G_0}  $. This is displayed in
Fig.~\ref{gmax}.  We take $\Re a > 0$ to ensure that
$\ee^{-G(V, {\bar V})}$ is positive in the
vicinity of $V =0$.

Observe that adding a cubic polynomial (and in particular a
linear term) in $V$ to (\ref{f0basic}) does
not affect the leading
behavior of (\ref{entrot}) near $V=0$.

Next, we compute the metric on the moduli space near $V =0$.  Using
 \begin{equation}
 g_{V \bar V} = \partial_V \partial_{\bar V} G =
 - \ee^G \, \partial_V \partial_{\bar V} \ee^{-G} + G_V \, G_{\bar V}
\spc,
  \end{equation}
and taking $V \rightarrow 0$, we find
$ G_V = - \ee^G \partial_V \ee^{-G}
\rightarrow 0$ and
 \begin{equation}
 g_{V \bar V} \approx \frac{\beta}{\pi} \, \ee^{G_0} \, \log\abs{V}^2
\spc.
 \label{metrict}
 \end{equation}
Note that the result (\ref{metrict}) depends crucially on having
$\Re a > 0$. Furthermore, for the metric to be positive
definite as $V \rightarrow 0$, the constant $\beta$ has to be
negative.

We also compute the gauge couplings associated with (\ref{f0basic}) near
$V \approx 0$.  Using \cite{deWit:1995zg}
\begin{equation}
g_{IJ}^{-2} = \frac{\im}{4} \left(\mathcal{N}_{IJ}
 -\bar{\mathcal{N}}_{IJ}\right),\qquad \mathcal{N}_{IJ} =
\bar F_{IJ} + 2\im \, \frac{\Im F_{IK} \Im F_{JL} Y^K Y^L}{\Im F_{MN}
Y^M Y^N}
\spc,
\label{gaugecoup}
\end{equation}
we find,
upon diagonalization, that one of the gauge couplings
remains approximately constant, while the other coupling
exhibits a logarithmic running,
\begin{equation}
g^{-2} \approx \Re a,  \;\;\spc,\;\;\;{\tilde g}^{-2}
\approx \frac{\beta}{4\pi}\log\abs{V}^2 \spc.
\label{gaugesugra}
\end{equation}
Observe that for negative $\beta$,
the coupling ${\tilde g}$ becomes small as $V \rightarrow 0$.

The basic example (\ref{f0basic}), with $\beta =- 1/2$,
describes the conifold singularity of the mirror quintic in type
IIB with $V = \psi -1$
(cf.~(\ref{f0f1con})) \cite{Candelas:1990rm,Ghoshal:1995wm}.
The metric (\ref{metrict}) is precisely the metric at the
conifold point $\psi =1$ (see table 2 of \cite{Candelas:1990rm}). Since
the
conifold singularity is associated with the appearance of one
additional massless hypermultiplet \cite{Strominger:1995cz}, we
see that we have entropy maximization when
a hypermultiplet becomes massless at the singularity.
Note that the character of the extremum of the entropy (\ref{entro2d})
at $V =0$ is independent of the value of $\Yf$.

Next, consider the resolved conifold in type IIA.  The associated
$\Fc^{(0)}$ is described by (\ref{f0basic}) with $\beta = - 1/2$
and $a = 0$ (cf.~(\ref{f0f1con})) \cite{Gopakumar:1998ki}.
We compute the K\"ahler metric and the
gauge coupling in the associated field theory. We decouple gravity
by restoring Planck's mass in (\ref{G2d})
with $G/M^2_{\mathrm{Pl}}$
and $z^A/M_{\mathrm{Pl}}$,
and by expanding
both sides of  (\ref{G2d})
in powers of $M^{-2}_{\mathrm{Pl}}$ \cite{Seiberg:1994rs},
\begin{equation}
G = {\hat K}(z, {\bar z})
+ f(z)  + {\bar f} ({\bar z})
+ \bigO(M^{-2}_{\mathrm{Pl}}) \spc,
\end{equation}
where
\begin{equation}
{\hat K} = - \left( {\bar z}^A \, \Fc_A + z^A {\bar \Fc}_A \right) .
\end{equation}
Using (\ref{f0basic}), we obtain near
$V = -\im z^1 \rightarrow 0$,
\begin{equation}
{\hat K} \approx \frac{ 2
\beta}{\pi} \abs{V}^2 \log\abs{V} \spc.
\end{equation}
Computing the corresponding K\"ahler metric near $V \rightarrow 0$
yields
\begin{equation}
g_{V {\bar V}} = \partial_V \partial_{\bar V} {\hat K} \approx \frac{ 2
\beta}{\pi} \log\abs{V} \spc,
\end{equation}
which is positive definite for
$\beta < 0$.
The gauge coupling is computed from (\ref{gaugecoup}) with
$\mathcal{N}_{IJ}
= {\bar F}_{IJ}$ \cite{deWit:1995ip}.  We obtain
\begin{equation}
{\tilde g}^{-2} = \frac{\im}{4} \left( {\bar F}_{11} - F_{11}
\right)
\approx \frac{\beta}{4\pi}\log\abs{V}^2 \spc,
\end{equation}
in agreement with (\ref{gaugesugra}).

More generally, whenever the singularity in moduli space is such that
a sufficiently large number
of (charged) hypermultiplets becomes massless there, so that the
resulting
$\beta$ is negative,
the function
(\ref{G2d}) exhibits a local maximum.
Examples thereof are
singularities
associated with the appearance of non-abelian gauge symmetries with
a non-asymptotically free spectrum
\cite{Klemm:1996kv,Katz:1996ht}.
A concrete example is provided by the so-called heterotic $S-T$ model,
which is
a two-K\"ahler moduli model with a dual type IIA description
in terms of a hypersurface of degree 12 in weighted projective space
$P^4_{(1,1,2,2,6)}$ with Euler characteristic
 $\chi = - 252$ \cite{Kachru:1995wm}.
The type IIA dual description is based
on (\ref{f02a}) with  $V = S -T$ and $n_{0,1} = 2$.  From (\ref{betan})
and (\ref{ac2}) we infer that $\beta = -1$ and that
$a$ is positive.  At $V=0$, a
gauge symmetry enhancement takes place,
whereby a $U(1)$ group is enlarged to
an $SU(2)$, with four additional (charged) hypermultiplets becoming
massless
there \cite{Klemm:1996kv,Katz:1996ht}.
The entropy of axion-free black holes in this model does indeed
have a maximum at $S=T$ (cf.~eq.~(4.34) in \cite{Behrndt:1997ei}).

\section{Entropy maximization in the presence of
$R^2$-interactions\label{emr2}}

Next, let us discuss entropy maximization in the presence of
higher-curvature
interactions encoded in  $F(Y, \Upsilon)$.  Usually,
the quantity $F(Y, \Upsilon)$ is assumed to have a
perturbative expansion of the form
\begin{equation}
F(Y, \Upsilon) = \sum_{g=0}^{\infty} F^{(g)} (Y) \, \Upsilon^{g} \spc.
\label{eq:expansionf}
\end{equation}
Then, expanding the entropy (\ref{entro}) in terms of the
coupling functions $F^{(g)} (Y)$ yields
\begin{equation}
\label{entrofg}
\begin{split}
\Sc/\pi &= \abs{Y^0}^2 \, \ee^{- G(z, {\bar z})} - 2\im \Upsilon
\left( F^{(1)} - {\bar F}^{(1)} \right) -\im \Upsilon (z^A - {\bar z}^A)
\left( \frac{{\bar Y}^0}{Y^0} \, \frac{\partial F^{(1)}}{\partial z^A}
 +  \frac{{Y}^0}{{\bar Y}^0} \, \frac{\partial {\bar F}^{(1)}}{\partial
{\bar z}^A} \right) \\
&\quad + 2\im \sum_{g =2}^{\infty} F^{(g)} (Y) \, \Upsilon^g \left( - g
+ (1-g) \, \frac{{\bar Y}^0}{Y^0} \right)
-
2\im \sum_{g =2}^{\infty} {\bar F}^{(g)} ({\bar Y}) \, \Upsilon^g \left(
- g
+ (1-g) \, \frac{{Y}^0}{{\bar Y}^0} \right) \\
& \quad -\im (z^A - {\bar z}^A) \sum_{g=2}^{\infty} \Upsilon^g
\left( \frac{{\bar Y}^0}{Y^0} \, \frac{\partial F^{(g)}}{\partial z^A}
 +  \frac{{Y}^0}{{\bar Y}^0} \, \frac{\partial {\bar F}^{(g)}}{\partial
{\bar z}^A} \right) ,
\raisetag{30pt}
\end{split}
\end{equation}
where $G(z, {\bar z})$ is given in (\ref{G2d}).

Let us first discuss the effect of the gravitational coupling function
$F^{(1)}$ on the maximization of the
entropy.  Let us again consider a singularity of the type
(\ref{f0basic})
associated with a vanishing
modulus $V= -\im z^1 = -\im Y^1/Y^0$,
while the other moduli are non-vanishing and kept fixed.
{From} (\ref{fupsexp}) it follows that $F^{(1)}$ takes the form
\begin{equation}
F^{(1)} \approx -
 \frac{\im}{64\cdot 12\pi} \beta \,\log V
\label{f1v0}
\end{equation}
near $V = 0$.
We compute (using $\Upsilon = - 64$)
\begin{equation}
- 2\im \Upsilon
\left( F^{(1)} - {\bar F}^{(1)} \right)=
4 \Im \left( \Upsilon \,
F^{(1)} \right) =
\frac{\beta}{3 \pi} \log\abs{V} \spc,
 \end{equation}
which for
negative $\beta$ reaches a maximum as $V \rightarrow 0$, i.e.~$\Im
\left(
\Upsilon \, F^{(1)} \right) \rightarrow + \infty$.

On the other hand,
the term proportional to $F^{(1)}_1$ in (\ref{entrofg})
only contributes with the phases of $Y^0$ and $Y^1$,
\begin{equation}
-\im \Upsilon (z^1 - {\bar z}^1)
\left( \frac{{\bar Y}^0}{Y^0} \,\frac{\partial F^{(1)}}{\partial z^1}
 +  \frac{{Y}^0}{{\bar Y}^0}\, \frac{\partial {\bar F}^{(1)}}{\partial
{\bar z}^1} \right)
= \frac{\beta}{6 \pi} \left( \cos (2 \theta_0) - \cos (2 \theta_1)
\right) ,
\end{equation}
where $Y^0 = \abs{Y^0} \, \exp(\im \theta_0)$
and $Y^1 = \abs{Y^1} \, \exp(\im \theta_1)$.

We conclude that, for negative $\beta$, not only does
$\ee^{-G}$ have a maximum at $V=0$, but also the
$R^2$-corrected entropy (\ref{entrofg}) based on $F^{(0)}$ and
$F^{(1)}$.
Moreover, the contribution of the coupling function $F^{(1)}$
is such that it enhances the maximization of the entropy.

The gravitational coupling function $F^{(1)}$ (as well
as the higher $F^{(g)}$) is known
to receive non-holomorphic corrections
\cite{Bershadsky:1993ta,Bershadsky:1993cx}.
For instance, for the quintic
threefold \cite{Bershadsky:1993ta} (and up to an overall constant),
\begin{equation}
\Re F^{(1)} = \log \left(  g^{-1}_{\psi {\bar \psi}} \,
\ee^{\frac{62}{3} K}
\, \abs{\psi^{\frac{62}{3}} \, ( 1 - \psi^5)^{- \frac{1}{6}}}^2 \right)
.
\end{equation}
Near $V = -\im z \equiv \psi - 1 \approx 0$, $K = \text{constant}$ and
$g_{\psi {\bar \psi}} \sim - \log\abs{V}$ \cite{Candelas:1990rm}, so
that
\begin{equation}
\Re F^{(1)} \sim - \log ( - \log\abs{V} ) - \ft16 \log\abs{V}^2 \spc.
\end{equation}
Therefore, as $V \rightarrow 0$, the behavior of the non-holomorphic
term is less singular than the behavior of the holomorphic term, and it
can be dropped from the maximization analysis.

Let us express $V = -\im Y^1/Y^0$ in terms of the charges
$q_0, q_1, p^0$ and $p^1$ by
solving the attractor equations (\ref{eq:attractor}).
We take $\Fc^{(0)}$ to be given by (\ref{f0basic}) and $F^{(1)}$ to be
given by (\ref{f1v0}).  For simplicity, we take
$Y^1$ to be imaginary and $Y^0$ to be real, so that $V$ is real.
Then
$Y^1 =\im \, p^1/2$, and $Y^0$ is determined by
\begin{equation}
4 (\Re a) (Y^0)^2 + q_0 Y^0 + \frac{\beta}{6 \pi} - \frac{\beta}{4 \pi}
(p^1)^2
= 0 \spc.
\end{equation}
A large value of $Y^0$ can be obtained by choosing $\abs{q_0}$ to be
large (assuming $\Re a \neq 0$),
whereas a small value of $V$ can be achieved by
sending $p^1 \rightarrow 0$.
Observe that taking $Y^0$ to be fixed at a large value is natural,
since (\ref{eq:expansionf}) is based on the
perturbative expansion of the topological string free energy,
and $Y^0$ is related to the inverse
 topological string coupling constant by
$Y^0= 2 \pi \,\gtop^{-1}$ (cf.~(\ref{dict})).

Next, let us discuss the effect of the higher $F^{(g)}$ (with $g \geq
2$)
on the maximization of the entropy.  It is known that the higher
$F^{(g)}$
also exhibit a singular behavior at $V = 0$.
For concreteness, we consider the conifold singularity of
the mirror quintic \cite{Candelas:1990rm}.
Near the conifold point $z= Y^1/Y^0 \rightarrow 0$
\cite{Bershadsky:1993cx,Ghoshal:1995wm},
\begin{equation}
F^{(g)} (Y) = \im
\frac{A_g}{(Y^0)^{2g-2} \, z^{2g-2} } \spc, \qquad g \geq 2 \spc,
\label{fgcon}
\end{equation}
where $A_g$ denote real constants which are expressed
\cite{Ghoshal:1995wm}
in terms of
the Bernoulli numbers $B_{2g}$ and are alternating in sign.
(cf.~(\ref{agrc})).

Inserting (\ref{fgcon}) into (\ref{entrofg}) yields
\begin{equation}
\label{entrofgz}
\begin{split}
\Sc/\pi &= \abs{Y^0}^2 \, \ee^{- G(z, {\bar z})} - 2\im \Upsilon
\left( F^{(1)} - {\bar F}^{(1)} \right) -\im \Upsilon (z - {\bar z})
\left( \frac{{\bar Y}^0}{Y^0} \, \frac{\partial F^{(1)}}{\partial z}
+  \frac{{Y}^0}{{\bar Y}^0} \, \frac{ \partial {\bar F}^{(1)}}{ \partial
{\bar z}} \right) \\
& \quad - 2  \sum_{g =2}^{\infty}
A_g \, \Upsilon^g  \, (Y^1)^{2-2g} \, \left( - g
+ (1-g) \, \frac{{\bar Y}^1}{Y^1} \right) \\
& \quad -2  \sum_{g =2}^{\infty} A_g \, \Upsilon^g \, ({\bar
Y}^1)^{2-2g}
\left( - g + (1-g) \, \frac{{Y}^1}{{\bar Y}^1} \right) .
\end{split}
\end{equation}
We observe that
the contribution from the higher $F^{(g)}$ ($g \geq 2$) to
(\ref{entrofgz})
does not cancel out.  This is problematic, since the
$F^{(g)}$ become increasingly singular as $z \rightarrow 0$.
For instance, when taking $Y^1$ to be purely imaginary,
the combination
$\Upsilon^g (Y^1)^{2 -2g}$ is negative for all $g \geq 2$,
and the total contribution of the higher $F^{(g)}$ to the entropy
does not have a definite sign due to the alternating sign of $B_{2g}$.
On the other hand, if we take $Y^1$ to be real, then the total
contribution of the higher $F^{(g)}$ to the entropy is positive, since
the combination
$A_g \Upsilon^g$ is positive for all $g \geq 2$.  This contribution
becomes
infinitely large at $Y^1 =0$.  Thus, to get a better
handle on the behavior of the entropy
in the presence of higher-curvature
corrections near the conifold point $V =0$, it is best to
use the non-perturbative expression for $F (Y, \Upsilon)$ for the
conifold
given in \cite{Gopakumar:1998ii,Gukov:2005bg}, rather than to rely on
the perturbative expansion (\ref{eq:expansionf}).
This will be done next.

For the resolved conifold in type IIA, $F(Y, \Upsilon)$ is given by
(see appendix A)
\cite{Gopakumar:1998ii,Gukov:2005bg},
\begin{equation}
F(Y, \Upsilon) = - C \,
\sum_{n=1}^{\infty} n \log \left(1 - q^n Q \right) ,
\label{resentro}
\end{equation}
where we neglected the $Q$-independent terms, since they do not affect
the extremization with respect to $Q$.
Here $q = \ee^{-\gtop}$ and $Q = \ee^{- 2 \pi V}$.
We impose the physical restrictions $\Re \gtop >0$ and $\Re V \geq 0$.
The topological string coupling constant $\gtop$ and the
constant $C$ are  expressed
in terms of $Y^0$ and of $\Upsilon$
via (\ref{dict}).

Under the assumption of uniform convergence,
inserting (\ref{resentro}) into
(\ref{entro})
yields the entropy
\begin{equation}
  \label{eq:SqQ}
  \Sc = -\sum _{n=1}^\infty \Re\left(
    \frac{n q^n Q \log q \bigl[n \log q + 2\log\abs{Q}\bigr]}
    {\log\bar{q} \, (1 - q^n Q) }
    +\frac{n \bigl[n q^n Q \log q - 2 (1 - q^n Q)\log(1-q^n Q)\bigr]}
    {1-q^n Q}\right).
\end{equation}
In order to determine the nature of the
extrema of (\ref{eq:SqQ}) we numerically approximated
this expression by a finite sum of a
sufficiently large number of terms. To improve the accuracy of the
approximation, the summation was performed in the order of decreasing
$n$, so that subsequent summands were of comparable magnitude to the
partial sums.

\begin{figure}
  \includegraphics[bb=91 3 322 190,width=.45\linewidth]{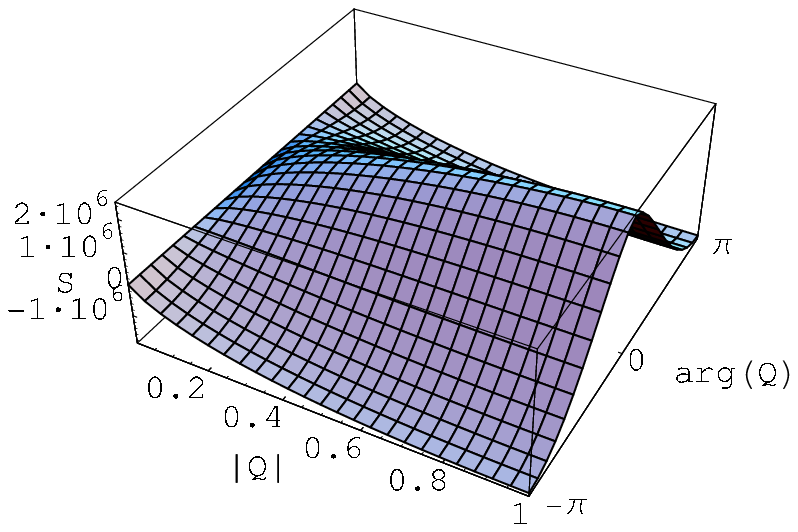}
  \hspace{\stretch{1}}
  \includegraphics[bb=91 3 322 190,width=.45\linewidth]{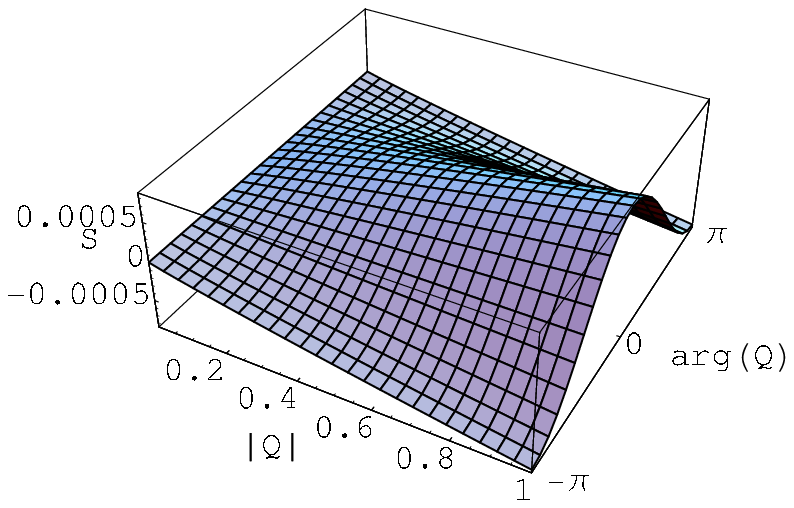}
  \caption{$\Sc$ as a function of $Q$ for $\gtop = 0.001$ (left) and
    $\gtop = 10$ (right). The conifold point corresponds to
    $\abs{Q} = \ee^{-\Re(2 \pi V)} = 1$ and $\arg(Q) = -\Im(2 \pi V) =
0$. The number of terms taken into account was 100001 and 1001,
respectively.}
  \label{fig:S_real_gtop}
\end{figure}

Let us first consider the case when $\gtop$ (and hence $q$)
is taken to be real.
We find that
the entropy attains a maximum at $Q = 1$, i.e.~at
the conifold point $V=0$, regardless of the
strength of the coupling $\gtop$, as
displayed in
Fig.~\ref{fig:S_real_gtop}. Observe that the maximum at $Q =1$ occurs
at the boundary of the allowed domain, where derivative tests do not
directly apply. $\Sc(q,Q)$ is periodic in $\arg(Q)=-\Im(2 \pi V)$.

\begin{figure}
  \centering
  \includegraphics[bb=91 3 322 146,width=.45\linewidth]{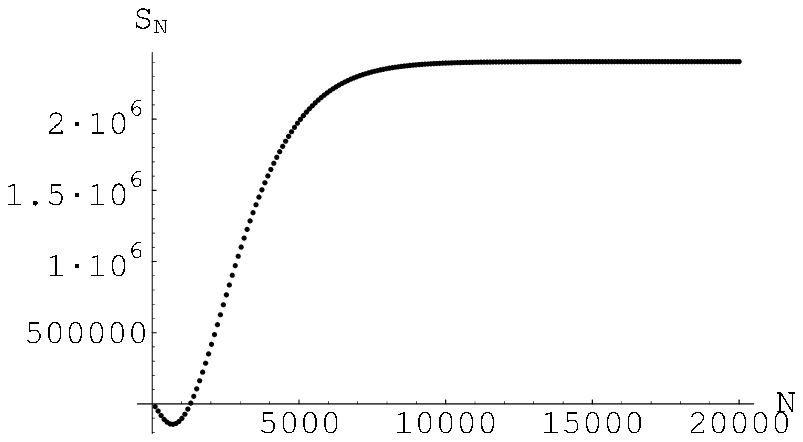}
  \hspace{\stretch{1}}
  \includegraphics[bb=91 3 322 146,width=.45\linewidth]{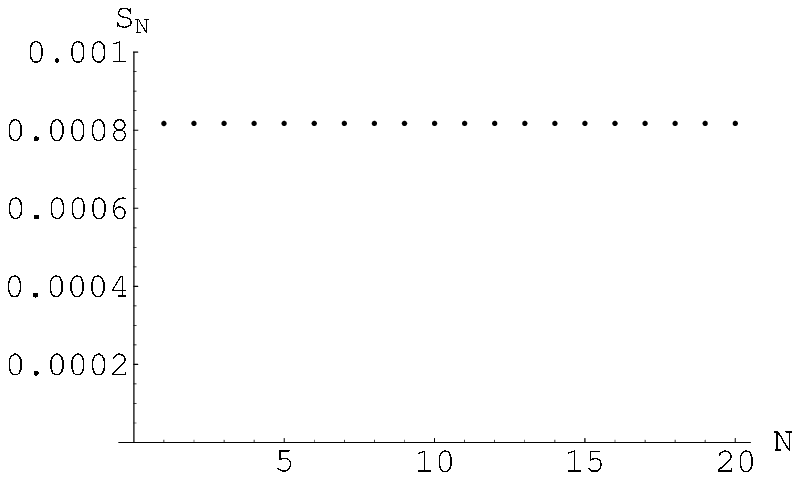}
  \caption{Sums of the first $N$ terms (partial sums) of
    \eqref{eq:SqQ} for $\gtop=0.001$ (left) and $\gtop=10$ (right) at
    the conifold point $V=0$ ($Q=1$).}
  \label{fig:S_convergence}
\end{figure}

As the string coupling becomes weaker,
the convergence of the series slows down, and the number of
terms needed to be taken into account grows roughly proportionally to
the inverse coupling, independently of $Q$.  This is
shown in Fig.~\ref{fig:S_convergence}.
We observe that apart
from the magnitude, the value of $\gtop$ has little influence on
the shape of $\Sc(q,Q)$.

\begin{figure}
 \includegraphics[bb=91 3 322 190,width=.45\linewidth]{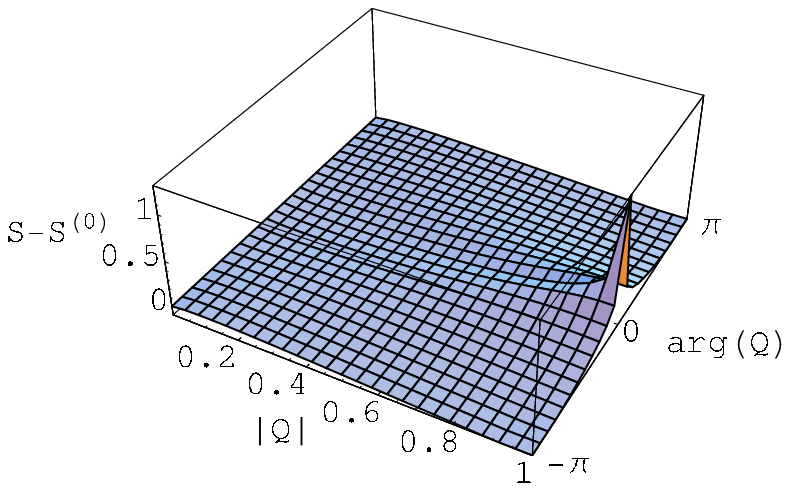}
 \hspace{\stretch{1}}
 \includegraphics[bb=91 3 322 190,width=.45\linewidth]{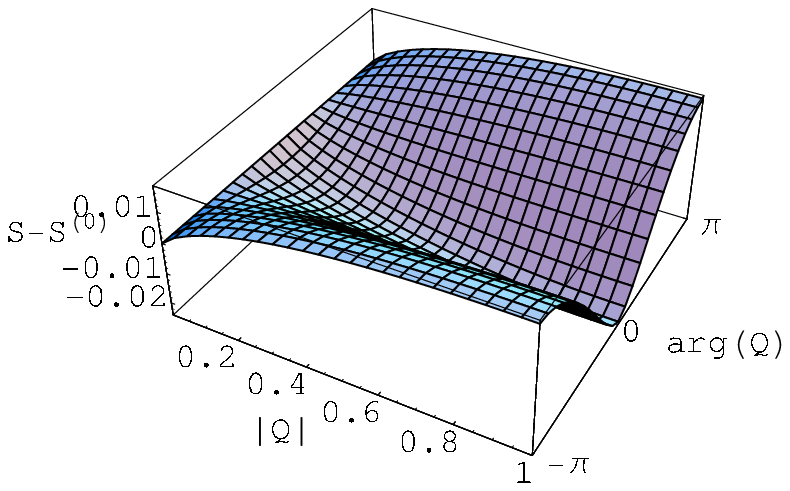}
 \caption{$\Sc-\Sc^{(0)}$ as a function of $Q$ for $\gtop=0.001$
 (left) and $\gtop=10$ (right).\label{fig:S-S0Qg}}
\end{figure}

Next, let us subtract the tree-level contribution to the entropy,
$\Sc^{(0)}$, computed from
\begin{equation}
F^{(0)} = C \, \gtop^{-2} \, \Li_3(Q) \spc,
\end{equation}
so as to exhibit the contribution to the entropy from the
higher-curvature
corrections.  The tree-level contribution
can be written as
\begin{equation}
  \Sc^{(0)}=\frac{2\Re\bigl[\Li_3(Q) - \log\abs{Q} \, \Li_2(Q)\bigr]}
             {\abs{\log q}^2} \spc.
\end{equation}
When treated as a function of $Q$ for a fixed $q$
(observe
that $\Sc^{(0)}$ does not depend on $\arg(\gtop)$),  $\Sc^{(0)}$
has a shape similar to the
shape of $\Sc$.

\begin{figure}
  \includegraphics[bb=91 3 322 190,width=.45\linewidth]{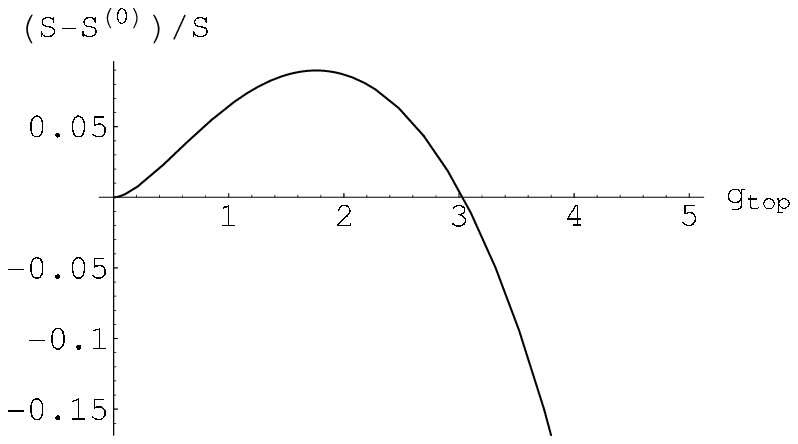}
  \hspace{\stretch{1}}
  \includegraphics[bb=91 3 322 146,width=.45\linewidth]{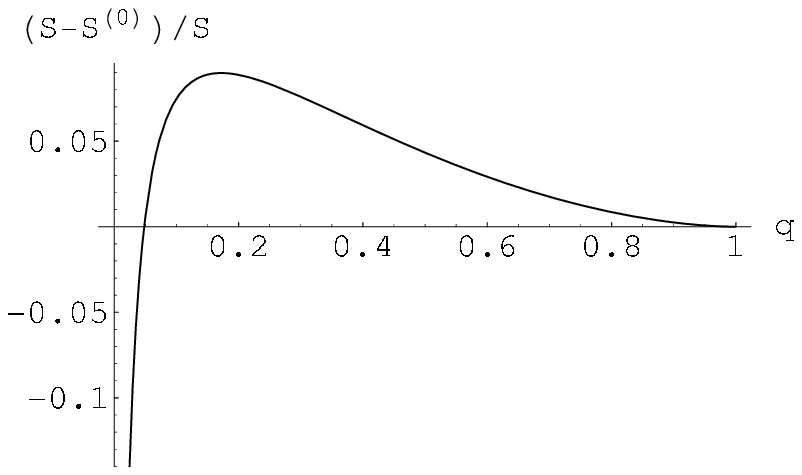}
  \caption{The ratio $(\Sc-\Sc^{(0)})/\Sc$ as a function of
  $\gtop$ or $q$ at $Q=1$.\label{fig:1-S0bySg}}
\end{figure}

The difference $\Sc-\Sc^{(0)}$ amounts to the contribution to the
entropy of
the higher-curvature
terms.  It depends on $Q$ and $\gtop$, as can be seen
in Fig.~\ref{fig:S-S0Qg}. At the conifold point $V=0$, the
difference $\Sc-\Sc^{(0)}$
is positive
for small coupling $\gtop$, whereas it becomes negative for large
values of $\gtop$.
At weak coupling higher-order corrections become negligible. At
strong coupling the corrections,
albeit smaller, are comparable to $\Sc^{(0)}$ and negative, resulting in
$\Sc\ll\Sc^{(0)}$ ($\Sc^{(0)}$ decreases as $\gtop^{-2}$, while $\Sc$
decreases as $\gtop\ee^{-\gtop}$ for large $\gtop$).
This is displayed in Fig.~\ref{fig:1-S0bySg}, where we plotted
$(\Sc-\Sc^{(0)})/\Sc$.

\begin{figure}
 \includegraphics[bb=91 3 322 190,width=.45\linewidth]{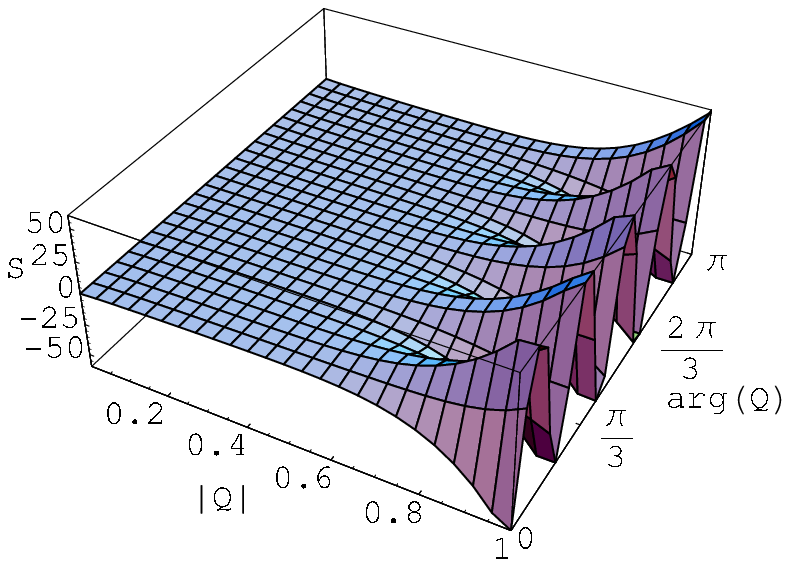}
 \hspace{\stretch{1}}
 \includegraphics[bb=91 3 322 190,width=.45\linewidth]{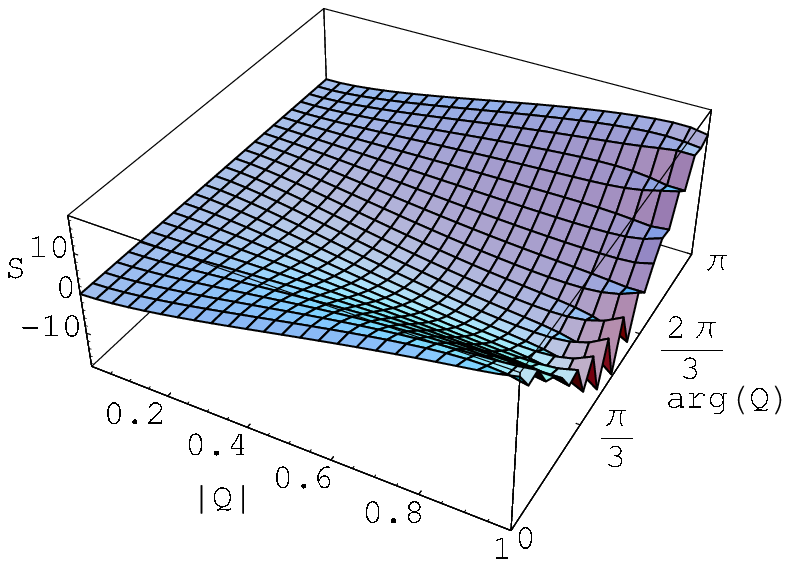}
 \\
 \includegraphics[bb=91 3 322 146,width=.45\linewidth]%
 {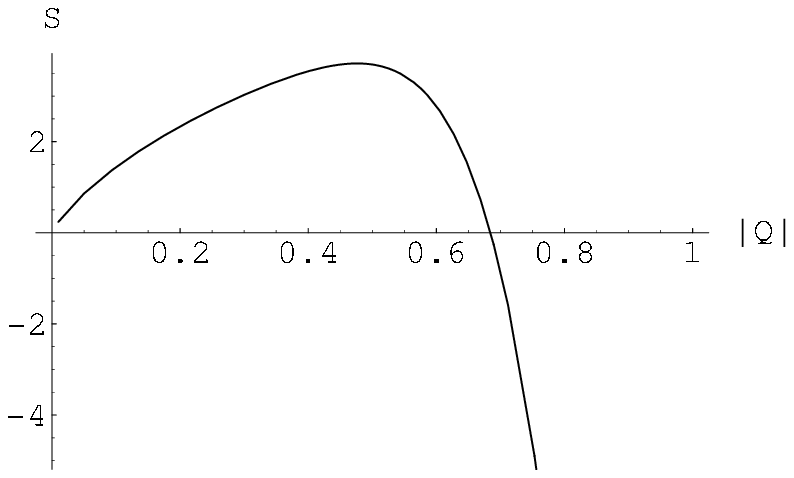}
 \hspace{\stretch{1}}
 \includegraphics[bb=91 3 322 146,width=.45\linewidth]%
 {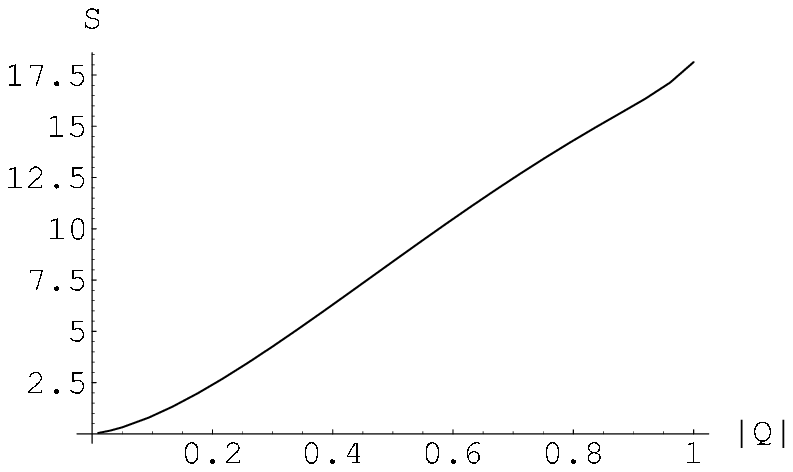}
 \\
 \includegraphics[bb=91 3 322 146,width=.45\linewidth]%
 {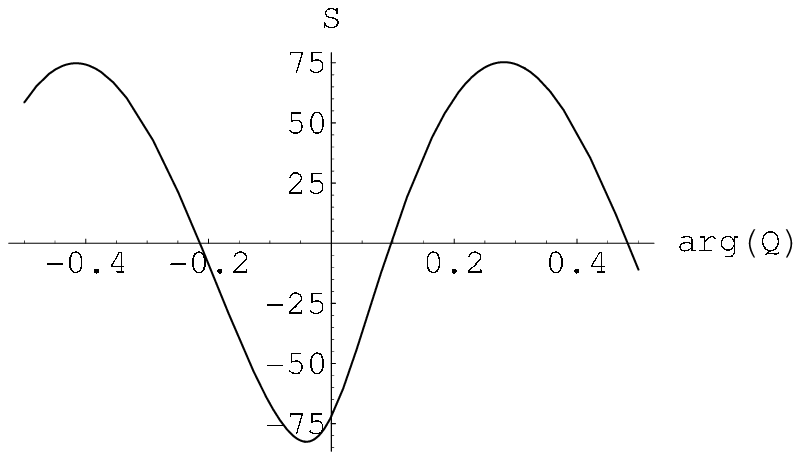}
 \hspace{\stretch{1}}
 \includegraphics[bb=91 3 322 146,width=.45\linewidth]%
 {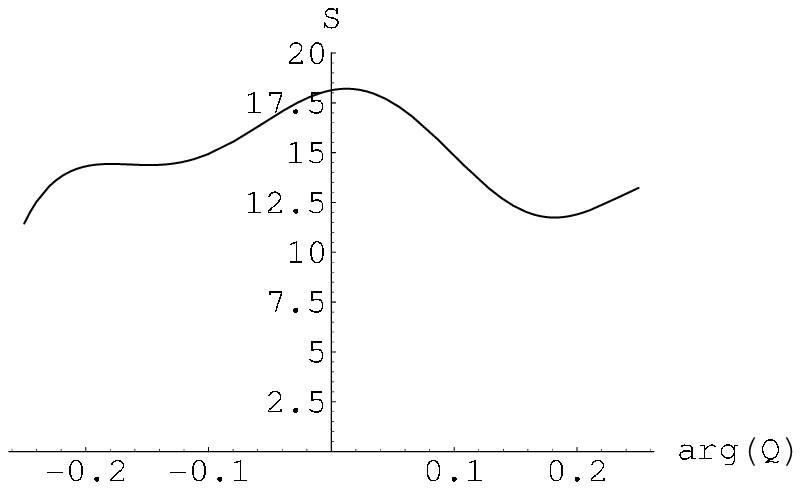}
 \caption{$\Sc$ as a function of $Q$ for complex $\gtop$. Left:
   $\gtop=0.01-0.7\im$, right: $\gtop=0.01+3\im$. Note that the
   range of $\arg(Q)$ in the 3-dimensional graphs has been cut by
   half to exhibit the point $Q=1$ more clearly (but the periodicity
   remains $2\pi$). The 2-dimensional graphs show in greater detail
   the edges of the surfaces closest to the viewer (cross-sections
   through the surfaces along $\arg(Q)=0$ and $\abs{Q}=1$).}
 \label{fig:S_cplx_gtop}
\end{figure}

Finally, if we allow $\gtop$ to be complex the
behavior of $\Sc$ changes markedly.  As  $\Re(V)$ tends to zero, we
notice increasingly pronounced oscillations, whose amplitude and
period sensitively depend on $\gtop$ (see Fig.~\ref{fig:S_cplx_gtop}).
In effect, the maximum formerly at $V=0$ is displaced and new local
extrema appear.

Note that the entropy (\ref{eq:SqQ}) is not necessarily positive,
because
we have not included the contribution stemming from
the Euler characteristic of the Calabi--Yau threefold (see (\ref{ac2}))
and of other moduli (which, if present, we have taken to be constant).

\section{Relation to OSV free energy \label{osvfree}}

According to \cite{Ooguri:2004zv},
the entropy (\ref{entro}) can be rewritten as
\begin{equation}
\Sc = E - L \spc,
\end{equation}
where $E$ denotes the OSV free energy which, in the conventions of
\cite{LopesCardoso:2006bg}, reads
\begin{equation}
\label{eq:OSV}
E = 4\pi\Im{F} \spc,
\end{equation}
and where $L$ is given by
\begin{equation}
L = \pi \, q_I \,\phi^I = 4 \pi \Im F_I \Re Y^I
\end{equation}
by virtue of the attractor equations
\begin{equation}
q_I = 4 \, \frac{\pd \Im F}{\pd \phi^I}
\end{equation}
with
$\phi^I = 2\Re Y^I$.

\begin{figure}[ht]
  \includegraphics[bb=91 3 322 190,width=.45\linewidth]{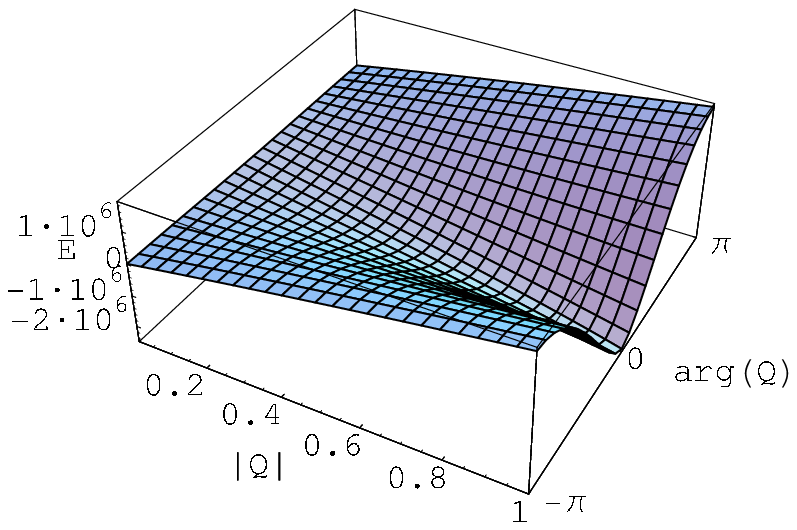}
  \hspace{\stretch{1}}
  \includegraphics[bb=91 3 322 190,width=.45\linewidth]{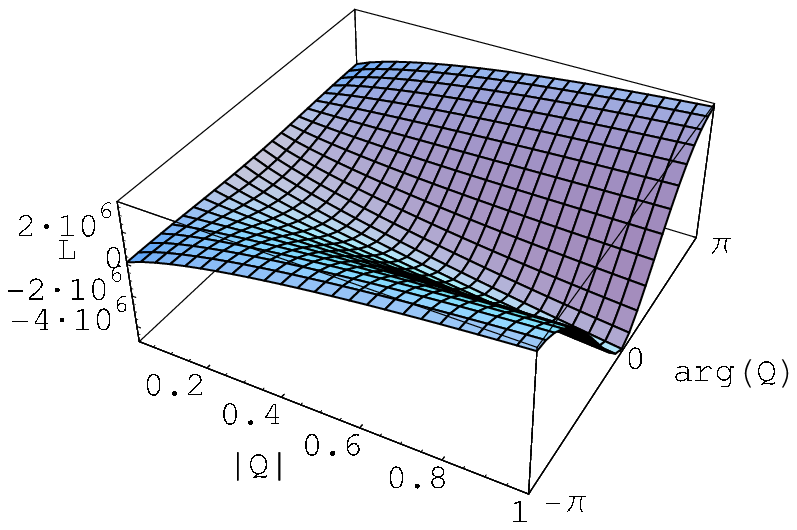}
  \caption{$E$ and $L$ as functions of $Q$ for $\gtop=0.001$
  (compare with the left graph in Fig.~\ref{fig:S_real_gtop}).}
  \label{fig:ELQ}
\end{figure}

The function $F(Y, \Upsilon)$ is related to the topological partition
function $\Ftop$ by
$F (Y, \Upsilon) = - \im \Ftop/(2 \pi)$ (see (\ref{fftop}) and
(\ref{dict}) with $\Upsilon = - 64$).  Using $E = -
(\Ftop + \bar{F}_{\mathrm{top}})$
and
\begin{equation}
\Ztop = \ee^{-\Ftop} \spc,
\end{equation}
we obtain
\begin{equation}
\label{eq:SandZtop}
\ee^\Sc = \abs{\Ztop}^2 \, \ee^{- L}\spc.
\end{equation}
For the resolved conifold in type IIA, the free energy $E$ and $L$,
computed
from (\ref{resentro}), are given by
\begin{equation}
\begin{split}
E &= 2 \sum_{n=1}^\infty n \Re\left(\log(1 - q^n Q)\right) \spc, \\
L &= 2 \sum_{n=1}^\infty n \left[\Im\left(\frac{q^n Q \log q}{1-q^n
Q}\right)
\Im\left(\frac{\log Q}{\log q}\right) \right.\\
&\qquad\qquad\; +\left.\Re\left(\frac{1}{\log q}\right)
\Re\left(\frac{q^n Q \log q (n \log q + \log Q)}{1-q^n Q} \right)
\right] .
\end{split}
\end{equation}
By numerically approximating these expressions as before, we find that
for
real coupling $\gtop$, the OSV
free energy $E$ is minimized at the conifold
point $Q=1$, see Fig.~\ref{fig:ELQ}. The entropy $\Sc$ is maximized at
the
conifold, as discussed before.

\begin{figure}
  \includegraphics[bb=91 3 322 146,width=.45\linewidth]{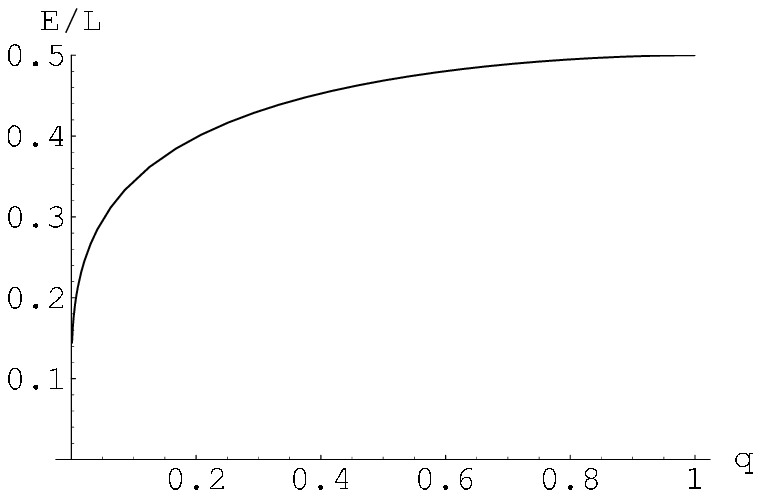}
  \hspace{\stretch{1}}
  \includegraphics[bb=91 3 322 146,width=.45\linewidth]{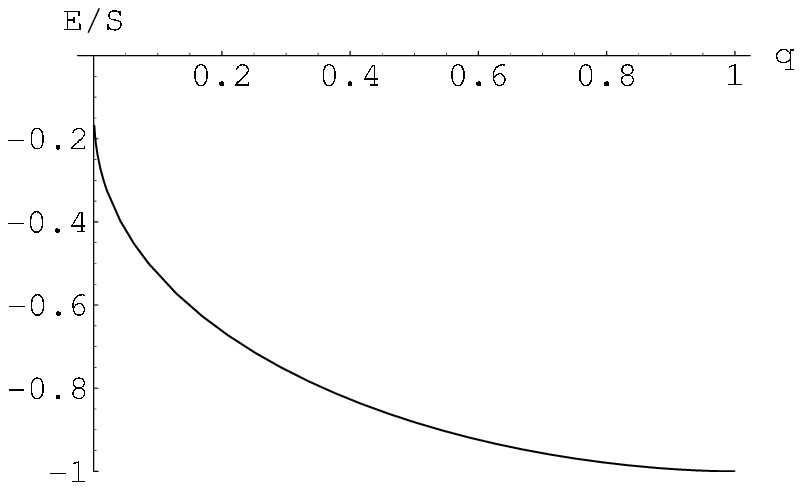}
  \caption{Ratios $E/L$ and $E/\Sc$ at
the conifold point, plotted as functions of real $q$.}
  \label{fig:ELq}
\end{figure}

At the conifold point, $E$ and $L$ as functions of real $q$ have the
behaviour displayed in Fig.~\ref{fig:ELq}.
In the limit $\gtop\rightarrow 0$, we thus find that at the conifold
point,
\begin{equation}
E = \frac12 L = - \Sc \spc
\end{equation}
(which holds for the sums, but not term by term).
Hence, at the conifold point,
\begin{equation}
\ee^{\Sc} =  \abs{\Ztop}^{-2} \spc.
\label{ztopcon}
\end{equation}
Observe that (\ref{ztopcon}) is in agreement with (\ref{eq:SandZtop})
at the conifold point.

The relation (\ref{ztopcon}) can also be derived from the prepotential
$\Fc^{(0)}$ given in (\ref{f0basic}), as follows.  At $V =0$,
by (\ref{entro2d}) and (\ref{entrot}),
\begin{equation}
\Sc = 2 \pi \, \abs{Y^0}^2 \, \left(\Fc^{(0)} + {\bar \Fc}^{(0)}\right)
.
\end{equation}
Taking $Y^0$ to be real (which corresponds to real $\gtop$),
and using $F = - \im (Y^0)^2 \, \Fc^{(0)}$,
we compute the OSV free energy (\ref{eq:OSV}) and find precisely
\begin{equation}
E = - \Sc \spc.
\end{equation}

\section{Conclusions\label{concl}}

We have discussed entropy maximization with respect to one complex
modulus
at points in moduli space
at which an excess of hypermultiplets becomes massless.
We found that the function $\ee^{- G(z, {\bar z})}$
exhibits a local maximum at such points.
When taking
into account the gravitational coupling $F^{(1)}$, the maximization
is further enhanced due to the additional term $\Im (\Upsilon
F_{\Upsilon})$ in the entropy (\ref{entro}) representing the departure
from the area law.  The inclusion of the higher $F^{(g)}$-couplings
into the maximization analysis is, however, problematic due to their
singular nature at these points in moduli space.  This problem can be
circumvented by resorting to the non-perturbative
expression for the topological
free energy, rather than relying
on its genus expansion.  We did so in the case of the resolved
conifold, where we found that the conifold point
is a maximum of the entropy for real topological string coupling, but
ceases to be a maximum once the topological string coupling is taken
to be complex.  Note that when
performing the maximization analysis we kept $Y^0$ fixed throughout
moduli space as in \cite{Gukov:2005bg}.  Other choices are, in
principle,
possible and could modify the results concerning
the maximization of the entropy.

\section*{Acknowledgements}

We would like to thank K.~Behrndt for collaboration at the early
stages of this work as well as for many valuable discussions. We
have also greatly benefited from discussions with G.~Curio, S.~Gukov,
P.~Mayr, K.~Saraikin and C.~Vafa.
This work is partly supported by EU contract MRTN-CT-2004-005104.

\appendix
\section{Normalization of
$F(Y,\Upsilon)$
\label{normalztn}}

In type IIA, $F^{(0)}(Y)$ has the following expansion
\cite{Candelas:1990rm,Candelas:1993dm,Hosono:1993qy,Hosono:1994ax}
\begin{equation}
F^{(0)}(Y) = (Y^0)^2 \left( - \frac{C_{ABC}  z^A z^B z^C}{6} + h(z) -
\frac{1}{(2 \pi \im)^3} \sum_{d_A} n_{d} \, \Li_3 ( \ee^{2 \pi \im
d_A z^A}) \right) ,
\end{equation}
where $C_{ABC}$ are the intersection
numbers and $n_d$ denote rational instanton numbers.  The quadratic
polynomial
$h(z)$ contains a constant term given by $\im \, \chi \,
\zeta (3)/(2 (2 \pi)^3)$, where $\chi$ denotes the Euler characteristic.
Using (\ref{f0f0}) yields
\begin{equation}
\Fc^{(0)}(z) = -\frac{\im}{6} C_{ABC} z^A z^B z^C +\im h(z) +
\frac{1}{(2 \pi )^3} \sum_{d_A} n_{d} \, \Li_3 ( \ee^{ 2 \pi\im \,d_A
z^A}) \spc.
\label{f02a}
\end{equation}
Observe that in the limit of large positive $\Im z^A$,
$\ee^{-G (z, {\bar z})}$  (computed from (\ref{G2d})) is positive,
as it should.

The coupling function $F^{(1)} (Y)$
is given by\footnote{We use the normalization given in
\cite{LopesCardoso:1998wt}.}
\cite{Bershadsky:1993ta,Candelas:1993dm,Hosono:1994ax}
\begin{equation}
{F}^{(1)} (Y) = - \frac{\im}{256 \pi}
\left[-
\frac{2 \pi \im}{12} c_{2A} \, z^A -  \sum_{d_A} \left( 2 n^{(1)}_d
\log ( \eta ( \ee^{2 \pi\im d_A z^A})) + \frac{n_d}{6} \log (1 -
\ee^{2 \pi\im d_A z^A}) \right) \right].
\end{equation}

Consider a singularity associated with the vanishing of one of
the moduli $T^A= -\im z^A$.  We denote this modulus by $V$. The other moduli
are taken to be large, so that we may approximate
\begin{equation}
\sum_{d_A} n_{d} \, \Li_3 ( \ee^{- 2 \pi d_A T^A}) \approx
\sum_{d_V} n_{0,0,\dotsc,d_V}  \, \Li_3 ( \ee^{- 2 \pi d_V V}) \spc.
\end{equation}
Let us assume that that the only non-vanishing instanton number $
n_{0,0,\dotsc,d_V} $ is the one with $d_V =1$. Using
\begin{equation}
\Li_3 (\ee^{-x}) = \zeta (3) - \frac{\pi^2}{6} x +
\left( \frac34 - \frac12 \log x \right) x^2 + \bigO(x^3) \spc,
\label{li3exp}
\end{equation}
we find that for $V \approx 0$, the function $\Fc^{(0)} $ can be
approximated by
\begin{equation}
\Fc^{(0)} = - \frac{C_{ABC} T^A T^B T^C}{6} +\im {\tilde h}(\im T)
+ \frac{\beta}{2 \pi} V^2 \log V  \spc,
\label{f0gen}
\end{equation}
where
\begin{equation}
\beta = - \frac{n_{0,0,\dotsc,1}}{2} \spc.
\label{betan}
\end{equation}
The instanton number $n_{0,0,\dotsc,1}$ counts the difference of
charged hyper- and vector multiplets becoming massless at $V=0$, i.e.
\begin{equation}
n_{0,0,\dotsc,1} = n_\mathrm{h} - n_\mathrm{v} \spc.
\end{equation}
Note that the quadratic polynomial $\im \tilde h$ contains a constant term
$a$ given by
\begin{equation}
a = (2 -\chi) \, \frac{\zeta (3)}{2 (2 \pi)^3} \spc.
\label{ac2}
\end{equation}
Similarly, we find that near $V =0$,
\begin{equation}
{F}^{(1)} (Y) =  - \frac{\im}{256 \pi} \left[
\frac{2 \pi }{12} c_{2A} \, T^A -  2 \sum_{d_A}
n^{(1)}_d \log ( \eta ( \ee^{- 2 \pi  d_A T^A})) +
\frac{\beta}{3} \log V  \right] .
\label{f1gen}
\end{equation}
Therefore, near $V = 0$ we obtain
\begin{equation}
\label{fupsexp}
\begin{split}
F(Y,\Upsilon) &= \sum_{g=0}^\infty F^{(g)}(Y) \Upsilon^g =
F^{(0)}(Y) + F^{(1)}(Y) \Upsilon + \dotsb \\
&= - \frac{\im (Y^0)^2}{2\pi} \beta \, V^2 \log V -
 \frac{\im\Upsilon}{64\cdot 12\pi} \beta \,\log V + \dotsb \spc,
\end{split}
\end{equation}
where we displayed only the terms proportional to $\log V$.

The function $F(Y, \Upsilon)$ is proportional to the topological
free energy $\Ftop(\gtop, z)$.
In order to determine the precise relation
between supergravity and topological string quantities, we consider the
case of the resolved conifold in type IIA.
First, observe that for this case
the functions $\Fc^{(0)}$ and $F^{(1)}$ are
given by \cite{Gopakumar:1998ki}
\begin{equation}
\begin{split}
\Fc^{(0)} &= - \frac{V^3}{12} +\im h(\im V) + \frac{1}{(2 \pi)^3} \sum_n
\frac{\ee^{- 2 \pi n V}}{n^3}  \spc, \\
{F}^{(1)} &= - \frac{\im}{256 \pi}\left[\frac{2 \pi}{12} \, c_2 \, V
- \frac{1}{6} \log (1-\ee^{- 2 \pi  V})\right] ,
\end{split}
\end{equation}
where $h(\im V)$ denotes a quadratic
polynomial in $V$, and where $c_2 = -1$. Observe that $\chi = 2$, so that
$a =0$.
Using (\ref{li3exp})
and (\ref{ac2}), it follows that near $V = 0$,
\begin{equation}
\label{f0f1con}
\begin{split}
\Fc^{(0)} &\approx - \frac{1}{4 \pi}
V^2 \log V  \spc, \\
{F}^{(1)} &\approx \frac{\im}{128 \cdot 12 \pi} \log V \spc.
\end{split}
\end{equation}
Then, comparison with (\ref{f0gen}) and
(\ref{f1gen}) yields $\beta = - 1/2$.

The topological
free energy for the resolved conifold
reads \cite{Gopakumar:1998ii,Gukov:2005bg}
\begin{equation}
\Ftop = -\sum_{n=1}^\infty n\log(1 - q^n Q) \spc,
\end{equation}
where $q = \ee^{-\gtop}$ and $Q = \ee^{-t}$, and
where we neglected the $Q$-independent terms.
We now review the standard argument leading to the expansion of
$\Ftop$ in powers of $\gtop$.
Using the Laurent
expansions
\begin{equation}
\log(1-z) = -\sum_{k=1}^\infty \frac{z^k}{k} \spc, \qquad \abs{z} < 1 \spc,
\end{equation}
and
\begin{equation}
\sum_{n=1}^\infty n z^n = \frac{z}{(1-z)^2} \spc, \qquad \abs{z} < 1 \spc,
\end{equation}
we obtain for $\abs{q^k Q} < 1$
and $\abs{q^k} < 1$,
\begin{equation}
\label{eq:Fexpansion}
\Ftop = \sum_{n=1}^\infty \sum_{k=1}^\infty \frac{n q^{kn} Q^k}{k}
= \sum_{k=1}^\infty \frac{q^k Q^k}{k (1-q^k)^2}
= \sum_{k=1}^\infty \frac{Q^k}{4k\sinh^2(k\gtop/2)} \spc.
\end{equation}
The conditions $\abs{q^k} < 1$ and $\abs{q^k Q} < 1$
imply that $\Re\gtop > 0$ and $\Re t > - \Re\gtop$,
the former condition being automatically satisfied for physical
coupling and the latter being fulfilled when $\Re t$
is interpreted as the volume of the two-cycle.

The expression (\ref{eq:Fexpansion})
can be further rewritten with the help of Bernoulli
numbers $B_n$, defined by
\begin{equation}
\label{eq:Bernoulli}
\frac{z}{\ee^z-1} = \sum_{n=0}^\infty B_n \frac{z^n}{n!} \spc, \qquad
  \abs{z} < 2 \pi \spc,
\end{equation}
and satisfying
\begin{equation}
B_{2n+1} = 0 \quad (n>0), \qquad B_{2n} = (-1)^{n-1}\abs{B_{2n}} \spc.
\end{equation}
The first few values are $B_0 = 1$, $B_1 = -1/2$,
$B_2 = 1/6$ and $B_4 = -1/30$.
Subtracting from \eqref{eq:Bernoulli} its
derivative multiplied by $z$ we obtain
\begin{equation}
\frac{(z/2)^2}{\sinh^2(z/2)} =
B_0 + \sum_{n=1}^\infty B_n
\frac{(1-n) z^n}{n!} \spc,
\end{equation}
and so, by virtue of the properties of $B_n$,
\begin{equation}
\Ftop = \sum_{k=1}^\infty \frac{Q^k}{k^3}\left(\gtop^{-2}
    + \sum_{g=1}^\infty (-1)^g \,
\frac{(2g-1)}{(2g)!}\abs{B_{2g}}k^{2g}\gtop^{2g-2} \right).
\end{equation}
This can be written in terms of the polylogarithms
\begin{equation}
\Li_s(z) = \sum_{k=1}^\infty \frac{z^k}{k^s} \spc,
\qquad \abs{z}<1 \spc,
\end{equation}
as
\begin{equation}
\Ftop = \gtop^{-2} \, \Li_3(\ee^{-t})
    + \sum_{g=1}^\infty (-1)^g
\, \frac{(2g-1)}{(2g)!} \, \abs{B_{2g}} \, \gtop^{2g-2} \, \Li_{3 -2g}
(\ee^{-t}) \spc.
\label{finftop}
\end{equation}
In the limit $ t \rightarrow 0$, we obtain \cite{Ghoshal:1995wm}
\begin{equation}
\begin{split}
\Ftop &= -\frac{1}{2}\left(\frac{t}{\gtop}\right)^2  \log t
+ \frac{1}{12}\log t - \sum_{g \geq 2}^{\infty} \frac{B_{2g}}{2g (2g-2)}
\left(\frac{\gtop}{t} \right)^{2g -2} \\
& \quad + \gtop^{-2} \, \zeta (3 ) + \sum_{g \geq 2}^{\infty} (-1)^g\,
\frac{(2g-1)}{(2g)!} \, \abs{B_{2g}} \, \gtop^{2g-2} \, \zeta (3 -2g) \spc,
\label{ffinexp}
\end{split}
\end{equation}
where we made use of the identity $\Li_s(1) = \zeta(s)$.

Observe that when deriving (\ref{finftop}) the expansion
(\ref{eq:Bernoulli}) was used, which is valid under the condition
$\abs{z}< 2 \pi$, or $\abs{k \gtop} < 2 \pi$.
In (\ref{eq:Fexpansion}), however,
this condition is satisfied only
up to a certain integer $k$.  The result (\ref{ffinexp}) is therefore
not rigorous.  A careful analysis of the asymptotic expansion at
weak topological coupling $\gtop$ has been given in
\cite{Dabholkar:2005by,Dabholkar:2005dt}.

Finally,
substituting $t = 2\pi V$ and comparing (\ref{ffinexp})
with (\ref{fupsexp}) yields
\begin{equation}
F(Y, \Upsilon)  = C F_{\mathrm{top}} (\gtop,z)
\label{fftop}
\end{equation}
with
\begin{equation}
\label{dict}
\begin{split}
C &= \frac{\im \, \Upsilon}{128\pi} \spc, \\
\gtop^2 &= -\frac{\pi^2 \Upsilon}{16 (Y^0)^2} \spc,
\end{split}
\end{equation}
where we used $\beta = - 1/2$.

For the conifold, it follows from (\ref{ffinexp}) and (\ref{dict}) that
in the limit $V = -\im Y^1/Y^0 \rightarrow 0$,
the higher coupling functions $F^{(g)} (Y)$ are given by
\begin{equation}
F^{(g)} (Y) =\im \frac{A_g}{(Y^1)^{2 g -2}} \spc, \qquad g\geq 2 \spc,
\end{equation}
where
\begin{equation}
A_g = - \frac{4^{2-2g}}{256 \pi} \, \frac{B_{2g}}{g (2g-2)} \spc.
\label{agrc}
\end{equation}
Observe that the coefficients $A_g$ are alternating in sign.


\addcontentsline{toc}{section}{\numberline{}References}
\bibliographystyle{JHEP-3}
\bibliography{bibliography}

\end{document}